\documentclass[aps,prb,twocolumn,groupedaddress,superscriptaddress,
amsfonts,amssymb,amsmath,a4paper]{revtex4-1}

\usepackage{amsmath,amssymb,graphicx,xcolor,bm}

\usepackage{hyperref}
\hypersetup{
	colorlinks,
	linkcolor={blue!75!black!80!yellow},
	citecolor={blue!75!black!80!yellow},
	urlcolor={blue!75!black!80!yellow}
}

\begin{document}

\title{Multipole plasmons and their disappearance in few-nanometer silver nanoparticles}
\date{\today}

\author{S\o ren Raza}
\affiliation{Department of Photonics Engineering, Technical University of Denmark, DK-2800 Kgs. Lyngby, Denmark}
\affiliation{Center for Nanostructured Graphene (CNG), Technical University of Denmark, DK-2800 Kgs. Lyngby, Denmark}
\affiliation{Department of Micro- and Nanotechnology, Technical University of Denmark, DK-2800 Kgs. Lyngby, Denmark}

\author{Shima Kadkhodazadeh}
\affiliation{Center for Electron Nanoscopy, Technical University of Denmark, DK-2800 Kgs. Lyngby, Denmark}

\author{Thomas Christensen}
\affiliation{Department of Photonics Engineering, Technical University of Denmark, DK-2800 Kgs. Lyngby, Denmark}
\affiliation{Center for Nanostructured Graphene (CNG), Technical University of Denmark, DK-2800 Kgs. Lyngby, Denmark}

\author{Marcel Di Vece}
\affiliation{Nanophotonics-—Physics of Devices, Debye Institute for Nanomaterials Science, Utrecht University, P.O. Box 80000, 3508 TA Utrecht, The Netherlands}

\author{Martijn Wubs}
\affiliation{Department of Photonics Engineering, Technical University of Denmark, DK-2800 Kgs. Lyngby, Denmark}
\affiliation{Center for Nanostructured Graphene (CNG), Technical University of Denmark, DK-2800 Kgs. Lyngby, Denmark}

\author{N. Asger Mortensen}
\affiliation{Department of Photonics Engineering, Technical University of Denmark, DK-2800 Kgs. Lyngby, Denmark}
\affiliation{Center for Nanostructured Graphene (CNG), Technical University of Denmark, DK-2800 Kgs. Lyngby, Denmark}

\author{Nicolas Stenger}
\email[E-mail: ]{niste@fotonik.dtu.dk}
\affiliation{Department of Photonics Engineering, Technical University of Denmark, DK-2800 Kgs. Lyngby, Denmark}
\affiliation{Center for Nanostructured Graphene (CNG), Technical University of Denmark, DK-2800 Kgs. Lyngby, Denmark}

\begin{abstract}
In electron energy-loss spectroscopy (EELS) of individual silver nanoparticles encapsulated in silicon nitride, we observe besides the usual dipole resonance an additional surface plasmon (SP) resonance corresponding to higher angular momenta. We even observe both resonances for nanoparticle radii as small as 4 nm, where previously only the dipole resonance was assumed to play a role.  Electron beams positioned outside of the particles mostly excite the dipole mode, but the higher-order resonance can even dominate the dipole peak when exciting at the particle surface, the usual choice for maximal EELS signal. This allows us to study the radius dependence of both resonances separately. For particles smaller than 4 nm, the higher-order SP mode disappears, in agreement with generalized nonlocal optical response (GNOR) theory, while the dipole resonance blueshift exceeds GNOR predictions. Unlike in optical spectra, multipole surface plasmons are important in EELS spectra even of ultra-small metallic nanoparticles.
\end{abstract}

\maketitle

\section{Introduction}
The optical properties of noble metal nanoparticles are dominated by their ability to support localized surface plasmon (SP) excitations, which can be described as the collective oscillation of the free-electron gas confined to the metal surface. Coupling between light and SPs gives rise to numerous interesting phenomena, such as squeezing light beyond the diffraction limit\cite{Gramotnev:2010} and large enhancements of the local electric field.\cite{Kelly:2003} These SP-induced effects lead to several technological applications, including e.g. improvement of absorption in solar cells,\cite{Atwater:2010} on-chip routing of electromagnetic energy,\cite{Zia:2006} and sensing of biological molecules.\cite{Anker:2008} The ability to probe the plasmonic response of individual nanoparticles is of crucial importance for designing novel structures that can harvest the full potential of plasmonics. One promise of plasmonics is that functional structures can be made much smaller than the wavelength of light, but subwavelength spatial information cannot be obtained with the usual diffraction-limited optical experiments. In contrast to light-based measurement techniques, electron energy-loss spectroscopy (EELS)\cite{Abajo:2010} performed in a transmission electron microscope (TEM) offers the ability to map the plasmonic resonances of sub-wavelength metal nanostructures\cite{Nelayah:2007,Bosman:2007} and in ultra-confined geometries\cite{Raza:2014a} due to the \AA{}ngstr\"{o}m spatial resolution of the TEM and tightly-confined electromagnetic field generated by the swiftly-moving electrons.\cite{Abajo:2010} In conjunction with an electron monochromator, an energy resolution down to a few hundred meV (and decreasing\cite{Krivanek:2014}) can be routinely achieved, allowing for high spatial- and spectral-resolution studies of individual plasmonic structures.

The localized SP resonance energies of metal nanoparticles can be selectively controlled, such as by changing the size, shape, material or the environment of the nanoparticle.\cite{Kelly:2003} However, even for a fixed system, a metal nanoparticle can according to classical electrodynamics show several localized SP resonances corresponding to excitations of different multipolar order. In particular, a spherical metal nanoparticle of radius $R$ with permittivity $\varepsilon(\omega) = \varepsilon_\text{core}(\omega) - \omega_\text{p}^2/\omega^2$ fully embedded in a dielectric medium $\varepsilon_\textsc{b}$ has the non-retarded resonance energies $\omega_l$ governed by\cite{Sernelius:2001}
\begin{equation}
	\omega_l = \frac{\omega_\text{p}}{\sqrt{\varepsilon_\text{core}(\omega_l) + \frac{l+1}{l}\varepsilon_\textsc{b}}}, \label{eq:res_freq_nr}
\end{equation}
where $\omega_\text{p}$ is the plasma frequency of the bulk metal, $\varepsilon_\text{core}(\omega)$ is the frequency-dependent response due to the bound charges of the metal, and, importantly, $l$ denotes the angular momentum of the SP mode, where $l=1$ is the dipole mode, $l=2$ is the quadrupole mode, and so on. Far-field measurement techniques based on plane-wave scattering mainly probe the dipole mode (see Supplementary Note 1), while near-field techniques such as EELS provide access to modes of larger angular momentum as well.\cite{Abajo:2010,Christensen:2014} This property has been advantageously used in many EELS experiments to map and study the multipolar plasmonic response of different metal nanostructures.\cite{Nelayah:2007,Bosman:2007,NGom:2008,Nelayah:2009,Schaffer:2009,Chu:2009,Koh:2011,Nicoletti:2011,Alber:2011,Guiton:2011,Rossouw:2011,Bigelow:2012,Nicoletti:2013,Rossouw:2013,Wiener:2013,Zhou:2014,Tan:2014,Martin:2014} However, study of the higher-order (HO) modes (i.e., $l>1$) in small noble-metal nanoparticles ($R<25$~nm) is hindered by interband transitions and the screening of bound charges [accounted for through $\varepsilon_\textrm{core}(\omega)$ in Eq.~\eqref{eq:res_freq_nr}], which strongly dampens the modes. Additionally, the energy spacing between subsequent $l$ modes is decreased compared to simple metals where $\varepsilon_\textrm{core}\approx 1$, making it difficult to spectrally resolve the modes. These issues can be overcome by encapsulating the nanoparticles in an insulating medium, which shifts the localized SP resonances to lower energies where interband transitions and losses are less pronounced.\cite{Kreibig:1987} EELS measurements on ensembles of encapsulated potassium\cite{vomFelde:1988} and silver nanoparticles\cite{Kreibig:1970} have shown the presence of HO modes, yet unambiguous measurements on the single-particle scale have to our knowledge not been achieved.

On the single-particle scale, several issues of ensemble measurements such as an inhomogeneous size distribution and the coupling between neighbouring particles are automatically prevented. In addition, deterministic studies of single particles constitute an ideal scenario for direct comparison with theoretical EELS simulations, allowing for thorough examination of electrodynamic theory on the few-nanometre scale, where the validity of the constitutive relations in classical theory is debated.\cite{Duan:2012,Scholl:2012} In previous studies, EELS measurements on small metal nanoparticles\cite{Kreibig:1970,vomFelde:1988,Ouyang:1992,Scholl:2012,Raza:2013} were compared with simulations based on optical scattering, thereby neglecting the importance of the position of the electron beam (i.e., impact parameter), and therefore the excitation of HO modes. Theoretical EELS studies on nanoparticles have shown that positioning the electron beam close to the nanoparticle surface more strongly excites SP modes of all angular momenta including HO modes,\cite{Abajo:1999,Abajo:2010,Christensen:2014} providing theoretical evidence that HO modes should be taken into account (through EELS simulations) in the analysis of EELS data on individual small nanoparticles.

\begin{figure}[!t]
	\centering
	\includegraphics[scale=1]{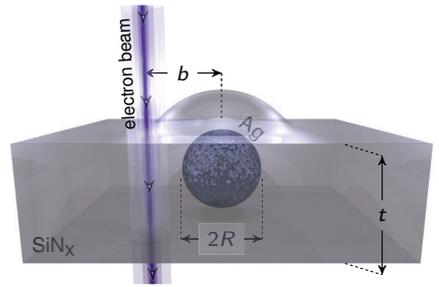}
	\caption{\textbf{Embedded silver nanoparticles.} Conceptual sketch of a silver (Ag) nanoparticle with radius $R$, which is embedded in a conformal silicon nitride (SiN$_\text{x}$) coating and excited by an electron beam. The impact parameter $b$ denotes the distance between the electron beam and the center of the nanoparticle. The thickness of the silicon nitride layer $t$ is determined from the EELS data and is approximately 35~nm, with about 20~nm below and 15~nm above the particles.}
	\label{fig:fig1}
\end{figure}

In this paper, we report EELS measurements on multi-twinned silver nanoparticles of icosahedral shape encapsulated in silicon nitride (see Fig.~\ref{fig:fig1} for illustration). We present experimental evidence of the excitation of both the dipole ($l=1$) and HO modes ($l>1$) in single silver nanoparticles in the radius range $1-20$~nm. Remarkably, we find that the HO modes can be identified down to a particle radius of only $4$~nm. To our knowledge, this is the first unambiguous experimental observation of HO modes in single silver nanoparticles of this ultrasmall size. We also provide experimental evidence of the significant dependence on the position of the electron beam when acquiring EELS spectra of small silver nanoparticles. Now, in previous studies it was reported that the relative weight of bulk and SP resonances can be controlled by the impact parameter.\cite{Scholl:2012} The experimental novelty here is that the impact parameter moreover also strongly influences the relative weight of the HO resonances. Even more surprisingly, for impact parameters close to the particle surface the spectral weight of the HO modes even exceeds that of the dipole mode. These results demonstrate that HO modes have unanticipated implications for the interpretation of EELS spectra of small nanoparticles.

Comparison with retarded EELS simulations based on classical electrodynamics, where we also take into account the energy resolution of the experimental setup, shows excellent agreement with measurements of nanoparticles down to $4$~nm in radius. Only for particle radii below $4$~nm do we observe a discrepancy between measurements and simulations, providing experimental evidence for a lower size limit for the application of classical electrodynamic theory. At this size scale the EELS signal from the HO modes disappears and we find only a single resonance, interpreted as the dipole mode, which strongly increases in energy with decreasing particle radius. We measure a blueshift of approximately 0.9~eV (from $2.8$~eV to $3.7$~eV) when the particle radius decreases from 4 to 1~nm, which cannot be explained by our simulations within classical electrodynamics. We explain the absence of HO resonances in the EELS signal as a consequence of size-dependent damping in small metal particles,\cite{Kreibig:1969} an effect recently linked to nonlocal response in metals by accounting for diffusion in the free-electron gas.\cite{Mortensen:2014}

\section{Results}
The TEM samples containing encapsulated silver nanoparticles are prepared by first depositing a thin layer of silicon nitride (approximately $15$~nm) on a commercially available silicon nitride TEM membrane ($5$~nm thickness). Subsequently, silver nanoparticles are deposited through sputtering (see Methods), ensuring a large size-distribution, which, crucially, allows all measurements to be performed under identical conditions on the same sample. Lastly, a final thin layer of silicon nitride (again, approximately 15~nm in thickness) is deposited to completely encapsulate the silver particles. The sample is maintained in vacuum during the entire fabrication process. By inspection in scanning TEM (STEM), we find that the silver nanoparticles are multi-twinned icosahedral in shape. The thickness of the silicon nitride layers is a trade-off between signal-to-noise ratio in the EELS measurements and the increase in permittivity of the background environment of the nanoparticles. Here, we find that a thickness of 15~nm provides a good compromise between encapsulation and membrane-induced loss. EELS measurements on a sample with thicker silicon nitride layers (approximately $30$~nm each, i.e., a total thickness of $t=60$~nm) showed a significant noise level in the EELS signal due to increased energy losses of the electrons in the thicker silicon nitride coating (data not included).

The morphology of the encapsulating silicon nitride layer was investigated by high-angular annular dark field (HAADF) STEM imaging on a TEM lamella prepared from a sample with thicker silicon nitride coating using focused ion beam milling. Complementing the cross-sectional view, we have also used plane-view imaging in HAADF STEM combined with energy-dispersive x-ray spectroscopy (EDS) and EELS to determine the morphology and composition of the sample presented in this work. Both of the different analyses reveal that the silicon nitride coating is conformal, i.e., the coating conforms to the topology of the nanoparticles as illustrated schematically in Fig.~\ref{fig:fig1}. The conformal coating ensures that particles of all sizes are being completely encapsulated (see Supplementary Note 3 for more details).

By encapsulating the silver nanoparticles in silicon nitride, several experimental key advantages are obtained as compared to previous STEM EELS studies of silver nanoparticles on substrates.\cite{Ouyang:1992,Scholl:2012,Raza:2013,Scholl:2013,Kadkhodazadeh:2013} Firstly, the encapsulated system ensures that the nanoparticles will not move along the surface (or even detach from the surface) due to the Coulomb forces from the nearby electron beam (which can occur for small metal particles on substrates\cite{Batson:2008,Batson:2011}). Secondly, the formation of silver sulfide and silver oxide is prevented. Finally, as also mentioned earlier, the large dielectric constant of the encapsulating silicon nitride redshifts the dipole and HO modes into a region of significantly lower loss of silver, allowing the probing of different SP modes almost separately in our EELS setup.

\begin{figure}[!b]
	\centering
	\includegraphics[scale=1]{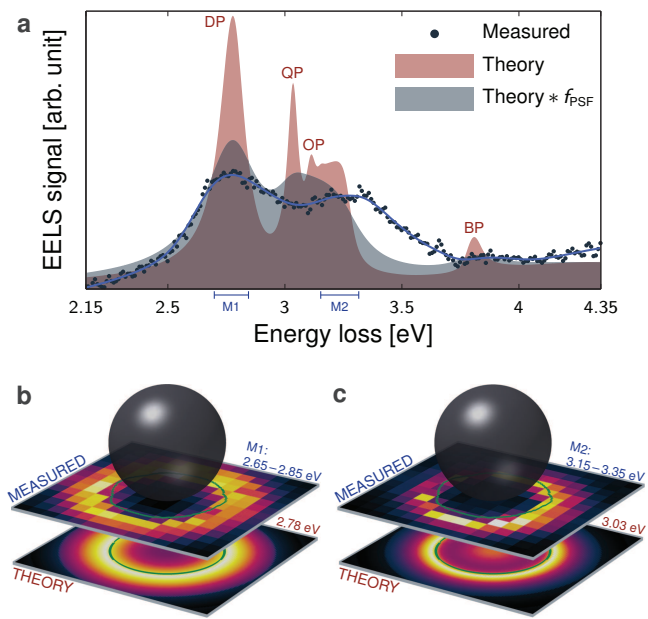}
	\caption{\textbf{EELS intensity maps of higher-order surface plasmons.} (\textbf{a}) Experimental EELS data as a function of energy loss of a SiN$_x$-encapsulated silver nanoparticle with radius $R=20 \pm 0.2$~nm excited by an electron beam positioned in the vicinity of the inner particle surface ($b=19.5 \pm 0.25$~nm). The red filled area shows the corresponding BEM EELS simulation, where the labels refer to dipole (DP), quadrupole (QP), octopole (OP), and bulk plasmon (BP). The gray filled area shows the theoretical curve (red filled area) convoluted with a Lorentzian point-spread function (PSF) with a FWHM of 0.15~eV ($f_\textsc{PSF}$). (\textbf{b-c}) Experimental (above), with zero-loss peak removed, and theoretical (below) EELS intensity maps of the same nanoparticle displaying the dipole and higher-order modes, respectively. The measured and simulated EELS intensity maps have identical color scales, respectively.}
	\label{fig:fig2}
\end{figure}

\begin{figure}[!b]
	\centering
	\includegraphics[scale=1]{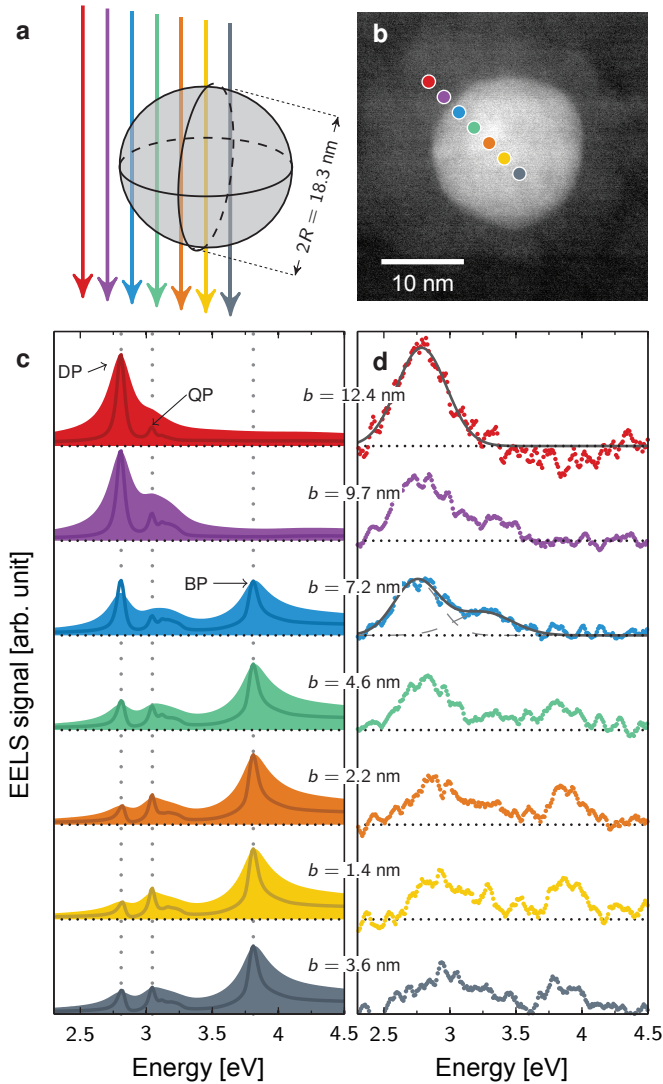}
	\caption{\textbf{EELS versus impact parameter.} (\textbf{a-b}) Theoretical setup and experimental STEM image, depicting also the impact parameters considered in the simulated and measured EELS spectra of (\textbf{c-d}), respectively, acquired from a silver nanoparticle with radius $R=9.2 \pm 0.2$~nm. The theoretical spectra (line plots) have been convoluted with a Lorentzian PSF with a FWHM of 0.15~eV (area plots). The convoluted spectra are normalized to unity area in the considered energy range. For clarity, the unconvoluted and convoluted spectra have the same maximal value. The measured spectra have been analyzed according to the procedure described in the Methods section and afterwards normalized to unity area in the same energy range. The fit of a single or two Gaussian functions to the peaks of interest in the experimental spectra is indicated in solid and dashed gray lines. In the measurements, the STEM probe diameter of $0.5$~nm is used as an estimate for the error in the impact parameter $b$.}	
	\label{fig:fig3}
\end{figure}

The EELS measurements are performed in a monochromated and aberation-corrected FEI Titan operated in STEM mode at an acceleration voltage of 120~kV, providing a probe size of approximately $0.5$~nm and an energy resolution of $0.15\pm0.05$~eV. In Fig.~\ref{fig:fig2}(a) we show the EELS spectrum at fixed impact for an encapsulated silver nanoparticle of radius $R=20\pm 0.2$~nm, which is the largest particle studied in this work. The post-processing routine of the EELS data along with the determination of the particle radius including its uncertainty are detailed in the Methods section. The impact parameter $b$, which defines the distance from the center of the particle to the position of the electron beam, is $b=19.5$~nm, i.e., only $0.5$~nm from the inner surface of the particle. Figure~\ref{fig:fig2}(a) shows two clear peaks at 2.74~eV and 3.25~eV due to the excitation of the dipole and HO modes of the silver nanoparticle, respectively. The corresponding measured EELS intensity maps are shown in Fig.~\ref{fig:fig2}(b) and (c). By comparison of the two maps, it is evident that the HO modes exhibit a significantly larger degree of surface localization than the dipole mode. The different degrees of spatial localization in the EELS intensity maps is of crucial importance, as it affords a degree of control over the excitation of specific SP modes through the careful variation of the impact parameter, as we discuss in more detail in relation with Fig.~\ref{fig:fig3}.

Accompanying the EELS measurements in Fig.~\ref{fig:fig2}, we also present simulated EELS spectra which have been computed using the boundary-element method (BEM),\cite{Hohenester:2012,Abajo:2002} see the Methods section and Supplementary Note 1 for details on the implementation. The icosahedral shape of the nanoparticles are modelled as spheres, an approximation which is justified by the fact that the optical response of these two shapes is very similar.\cite{Noguez:2007,Yang:2011} We model the precise geometry of silver nanoparticles conformally encapsulated in finite-thickness silicon nitride layers, and use a permittivity of $\varepsilon_{\text{SiN}_x}=3.2$ for the silicon nitride. The Kramers--Kronig procedure\cite{Egerton:2011} did not give a reliable result for the dielectric function of the silicon nitride layer, see Supplementary Note 2 for details. We have therefore chosen a pragmatic approach to determining $\varepsilon_{\text{SiN}_x}$ by achieving the best correspondence between the simulated (classical electrodynamics) and measured dipole resonance energies for all of the particle sizes studied, except for the smallest particles ($R<4$~nm) where a blueshift of the dipole mode is observed. As we will see, this very same value for the permittivity of the silicon nitride ($\varepsilon_{\text{SiN}_x}=3.2$) also captures the HO resonance energies for all of the particle sizes examined. Additionally, the same value for the permittivity describes other EELS measurements made on \emph{gold} nanoparticles encapsulated in silicon nitride (data not included).

The red curve in Fig.~\ref{fig:fig2}(a) shows the result of the simulated EELS spectrum for the same particle radius ($R=20$~nm) and impact parameter ($b=19.5$~nm) as in the measurement. Clear spectrally separated peaks due to the excitation of the dipole, quadrupole, and octopole SP modes along with the bulk plasmon are present. We see that the dipole peak is captured in the EELS measurement, while the individual HO modes are not spectrally separated in the measurement. Instead the HO modes accumulate into a single broad peak due to the energy resolution of the EELS setup. To mimic the experimental energy resolution, we convolute our EELS simulation with a Lorentzian point-spread function (PSF) with a full-width at half maximum (FWHM) of 0.15~eV [grey curve in Fig.~\ref{fig:fig2}(a)]. In the convoluted case, the HO modes merge into a single broader peak with a resonance energy that is only slightly lower than in the EELS measurement. This small discrepancy may be due to the faceted nature of the nanoparticles, which may influence the resonance energy of the HO mode as the HO mode is more localized to the surface (and thereby more sensitive to the particular surface morphology) than the dipole mode. An additional feature of the convolution is the significant decrease in intensity of the bulk plasmon, which we also observe in the measured EELS spectrum of Fig.~\ref{fig:fig2}(a). Finally, Fig.~\ref{fig:fig2}(b) and (c) display theoretical EELS maps of the dipole and quadrupole SP modes, respectively, which confirms the experimentally observed stronger surface localization of the HO modes as compared to the dipole mode. In the following, we will discuss how we utilize this SP mode localization to probe the resonance energies of the dipole and HO modes separately. We start by considering the implications of the impact parameter on the EELS signal from a small silver nanoparticle (i.e., varying the impact parameter $b$ for a fixed particle radius $R$). Hereafter, we consider nanoparticles of different sizes with the electron beam close to the particle surface (i.e., varying the particle radius $R$ for almost fixed distance between impact parameters and particle surface $b-R$).

The relative excitation amplitudes of dipole, HO, and bulk plasmon modes depend strongly on the impact parameter $b$, i.e., the position of the electron beam when acquiring EELS data. This dependence is important to understand, since the resonance energies of the SP modes are spectrally close. The ideal situation is to find impact parameters where we primarily excite the dipole mode with a weak signal from the HO modes and vice versa. While challenging, this is to some extent possible: the dipole mode is primarily excited for the electron beam focused several nanometres outside the nanoparticle ($b>R$), while all SP modes including the HO modes are efficiently excited for the electron beam close to the surface of the particle ($b\approx R$). For electron-beam excitations inside the particle ($b<R$), the bulk plasmon is also excited. In Fig.~\ref{fig:fig3} we show experimental and simulated EELS spectra of an encapsulated silver nanoparticle with a radius of $R=9.2\pm0.2$~nm as a function of the impact parameter $b$. Figures~\ref{fig:fig3}(a) and (b) show a schematic and an experimental STEM image of the nanoparticle, respectively, along with the positions of the electron beam. The corresponding theoretical and experimental EELS spectra are shown in Fig.~\ref{fig:fig3}(c) and (d). The impact parameter is varied systematically through a radial progression from particle center to several nanometers outside its surface. In Fig.~\ref{fig:fig3}(c), the unconvoluted theoretical EELS spectra are shown as line plots, while the PSF-convoluted spectra are shown as filled area plots. Considering first the spectra for the largest impact parameters (red curves), it is clear from Fig.~\ref{fig:fig3}(c) that the simulated EELS signal primarily stems from the dipole mode. In the EELS measurement a single symmetric peak is observed, which we accordingly interpret as the dipole mode. A Gaussian function [black solid line in top panel of Fig.~\ref{fig:fig3}(d)] was fitted to determine its resonance energy. As the electron beam approaches the surface of the nanoparticle, the simulated spectra show that HO modes increasingly contribute to the EELS signal (purple and blue curves). This feature is also observed in our EELS measurements (purple curve), where we observe that the peak is no longer symmetric due to the EELS signal from the HO modes. For $b=7.2$~nm (blue curve) an additional peak (besides the dipole peak) shows up. By fitting the sum of two Gaussian functions [grey dashed lines in third panel from top in Fig.~\ref{fig:fig3}(d)] we determine the resonance energies of both of the peaks as the center of each Gaussian. The resonance energy of the low-energy peak is in good agreement with that measured for the dipole resonance energy for impact parameter outside the particle ($b>R$), and the high-energy peak is then attributed to the accumulated contributions from HO modes. As the electron beam penetrates closer to the center of the particle, the simulated EELS signal from the bulk plasmon increases significantly while the signal from the SP modes decreases. In particular the dipole mode decreases strongly in signal since the penetrating electron beam cannot properly induce the antisymmetric dipolar charge distribution in the particle. Briefly summing up, we have shown that the dipole mode can be probed selectively by positioning the electron beam outside the particle ($b>R$), while the signal from the HO modes can be maximized by considering impact parameters close to the surface of the particle ($b\approx R$).

\begin{figure}[!b]
	\centering
	\includegraphics[scale=1]{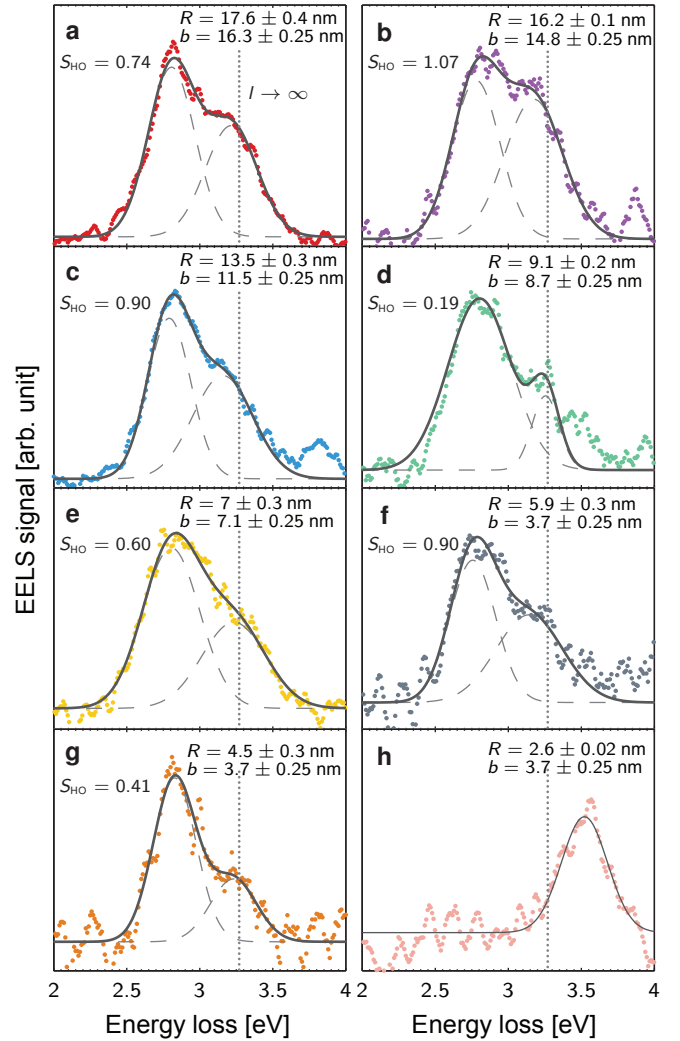}
	\caption{\textbf{EELS versus particle sizes.} (\textbf{a}-\textbf{g}) EELS spectra of silver nanoparticles encapsulated in silicon nitride used for determining the resonance energy of the HO surface plasmon modes. The least-squared fit of the sum of two Gaussian functions is indicated in solid gray lines, with each Gaussian indicated in dashed gray lines. The parameter $S_\textsc{ho}$ denotes the spectral weight of the HO modes, calculated as the ratio of the areas of the HO and dipole Gaussian functions. The dotted lines indicate the classical $l\rightarrow\infty$ energy limit. (\textbf{h}) EELS spectrum of encapsulated small silver nanoparticle (typical for all spectra for $R < 4$~nm) showing only a single blueshifted plasmon resonance, attributed to the dipole mode. The text inset of each plot designates the particle radius $R$ and the impact parameter $b$.}
	\label{fig:fig4}
\end{figure}

Next, we study the dipole and HO mode resonance energies of encapsulated silver nanoparticles as a function of particle radius $R$, while keeping the electron beam positioned close to the particle surface. To determine the \emph{dipole} resonance energy, we consider only impact parameters several nanometers outside the nanoparticle. In this case, the EELS spectra show a single symmetric peak [as that shown in the red curve of Fig.~\ref{fig:fig3}(d)], which we fit with a single Gaussian function to determine the resonance energy. However, the resonance energy of the \emph{HO modes} is determined by considering EELS spectra acquired by positioning the electron beam close to the surface of the nanoparticle. In these measurements as shown in Fig.~\ref{fig:fig4}(a-g), we consistently observe two peaks on silver nanoparticles of various radii (for fixed distance between electron beam and particle surface $b-R$), in agreement with what we found in the blue spectrum in Fig.~\ref{fig:fig3}(d) where the electron beam was also close to the surface. We stress that the clear and repeated observations in Fig.~\ref{fig:fig4}(a-g) of two SP peaks rather than one in various nanoparticles, and the low noise level, are aided by the fact that the strongest EELS signal is observed when exciting particles close to their surfaces. While individual silver nanoparticles have been studied before with EELS,\cite{Ouyang:1992,Scholl:2012,Raza:2013} the measurements presented here constitute, to our knowledge, the first unambiguous observation of HO modes on this ultra-small size and their controlled excitation.

\begin{figure}[!t]
	\centering
	\includegraphics[scale=1]{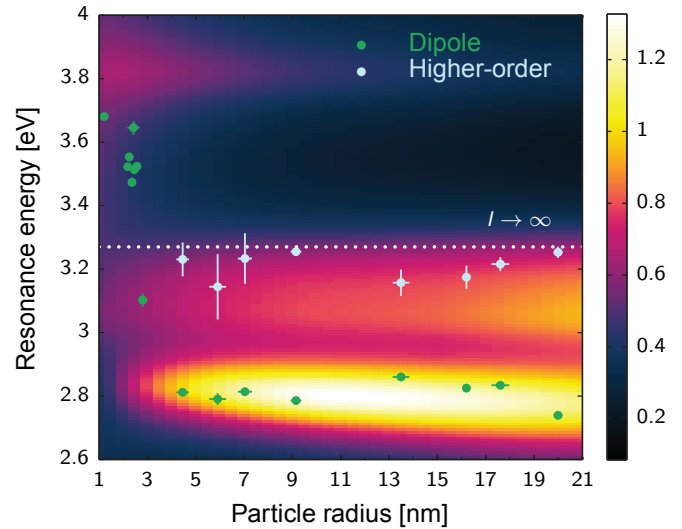}
	\caption{\textbf{Dipole and higher-order mode resonance energies.} Measured resonance energies as a function of particle radius for the dipole (green dots) and higher-order (light blue dots) surface plasmon modes in silver nanoparticles encapsulated in silicon nitride. The color plot shows the PSF-convoluted EELS simulations. For each particle size in the simulations, the electron beam is positioned 1~nm from the inner particle surface (i.e., $b-R=-1$~nm). Each simulated EELS spectrum has been normalized to unity area.}
	\label{fig:fig5}
\end{figure}

For every panel in Fig.~\ref{fig:fig4}(a-g) we determine the resonance energy of the HO modes by fitting the spectrum to the sum of two Gaussians (one for the dipole peak and one for the HO peak) and extracting the peak position of the Gaussian function fitted to the HO peak. The two-Gaussian fits accurately capture the features of the EELS data (see also Supplementary Figure~S1 for a residual plot of Fig.~\ref{fig:fig4}), which provides support that the HO modes peak can to a good approximation be described by a single (symmetric) Gaussian function. The reason for this is two-fold. Firstly, the individual HO modes are only closely spaced in energy (with decreasing energy spacing for increasing angular momentum $l$) as a consequence of the maximal allowed resonance energy of the HO modes in classical electrodynamic theory, given by the $l\rightarrow \infty$ limit of Eq.~\eqref{eq:res_freq_nr}. Secondly, the experimental PSF is a spectrally broad (compared to the individual HO modes) and symmetric function, which merges the individual spectrally-close HO modes into a single almost symmetric peak [see also Fig.~\ref{fig:fig2}(a)]. We can provide an accurate estimate for the energy of the $l\rightarrow \infty$ limit, which provides an upper energy limit for the HO modes, by considering the classical $l$-dependent resonance condition for a spherical particle in a homogeneous background medium, governed by\cite{Christensen:2014}
\begin{equation}
	l\varepsilon(\omega) + (l+1) \varepsilon_\textsc{b} = 0, \label{eq:res_cond}
\end{equation}
from which Eq.~\eqref{eq:res_freq_nr} is derived. The nonretarded approximation used in the derivation of Eq.~\eqref{eq:res_cond} is valid and accurate for silver nanoparticles with radius below 10~nm. To determine the energy of the $l\rightarrow\infty$ limit of Eq.~\eqref{eq:res_cond}, we require first an effective value for the background permittivity $\varepsilon_\textsc{b}$. We estimate the effective background permittivity by rewriting Eq.~\eqref{eq:res_cond} in the case of the dipole mode ($l=1$) as $\varepsilon_\textsc{b} = -\textrm{Re}[\varepsilon(\omega=\omega_\textsc{dp})] / 2$, where $\hbar\omega_\textsc{dp}$ is the resonance energy of the dipole mode. Here, we use the average value for the dipole resonance energies measured from nanoparticles in the radius range $4~\textrm{nm}<R<10~\textrm{nm}$ (Fig.~\ref{fig:fig5}), where the lower radius limit is needed since we measure an increase in the resonance energy for particles smaller than $R<4$~nm (we discuss this observation in more detail in relation to Fig.~\ref{fig:fig5}). From this procedure we find $\varepsilon_\textsc{b} = 3.3$, which is very close to $\varepsilon_{\text{SiN}_x}=3.2$ used for the silicon nitride layers in our BEM simulations, supporting also the interpretation that the smallest nanoparticles ($R<10$~nm) behave as if in a fully embedded homogeneous background environment. With $\varepsilon_\textsc{b} = 3.3$ as the value for the background permittivity, we find from Eq.~\eqref{eq:res_cond} the energy for the $l\rightarrow\infty$ limit to be $\hbar \omega_{l \rightarrow \infty} = 3.27$~eV, which, in the framework of classical electrodynamic theory, provides an upper limit for the resonance energy of the HO modes. Interestingly, the experimental EELS spectra shown in Fig.~\ref{fig:fig4}(a-g) (and the corresponding resonance energies shown in Fig.~\ref{fig:fig5}) indicate that the HO modes are bounded by this classical upper energy limit (shown as dotted lines in Figs.~\ref{fig:fig4} and \ref{fig:fig5}), since we do not measure any of the HO peaks to have a resonance energy larger than $\hbar \omega_{l \rightarrow \infty}$, although we do find some of the resonance energies to be very close to the limit.

As a quantitative measure of the strength of the HO modes, we additionally determine the relative spectral weight of the HO modes $S_\textsc{ho}$ as the ratio of the areas of the HO Gaussian fit to the dipole Gaussian fit. The values for $S_\textsc{ho}$ are shown in each panel in Fig.~\ref{fig:fig4}(a-g). We see that the relative spectral weight of the HO modes often exceeds 50\% and in one instance even surpasses 100\% [Fig.~\ref{fig:fig4}(b)], indicating that the EELS signal from the HO modes dominates that of the dipole mode. These large values for $S_\textsc{ho}$ suggest that EELS data acquired by positioning the electron beam at the surface of the nanoparticle cannot simply be interpreted solely as the dipole mode,\cite{Ouyang:1992,Scholl:2012,Raza:2013} but require careful consideration of the HO modes. 

The experimentally-observed dipole and HO mode resonance energies of all of the nanoparticles studied in this work are summarized in Fig.~\ref{fig:fig5}. Additionally, the superimposed color plot shows the simulated EELS spectra, where an impact parameter set to 1~nm from the inner particle surface is chosen (i.e., $b-R=-1$~nm) to strongly excite the HO modes (as also done in the EELS measurements). The simulated EELS spectra have been convoluted with a Lorentzian PSF and show a strong low-energy peak due to the dipole mode and an only slightly weaker high-energy peak due to the excitation of HO modes. Furthermore, an EELS signal from the bulk plasmon with an energy of approximately 3.8~eV is also present and increases in strength for decreasing particle size (since the impact parameter gets closer to the center of the particle). In the simulations, the dipole mode redshifts slightly with increasing particle radius due to retardation effects, while the HO modes remain almost constant in energy. For particle radii above 4~nm, we find excellent agreement between the observed and simulated resonance energies for both the dipole and HO modes, thereby remarkably validating classical electromagnetic theory down to the few-nanometer scale.

For particle radii below 4~nm, we experimentally observe two distinct features that are not explained by our classical simulations. Firstly, the EELS signal from the HO mode decreases significantly (and basically disappears), since we only observe one single peak in all of our EELS measurements regardless of the impact parameter. We interpret this peak as the dipole mode, since a strong EELS signal is always present for impact parameters outside the nanoparticles, see Fig.~\ref{fig:fig4}(h) for a typical example of such an EELS spectrum. Figure~\ref{fig:fig5} shows that the EELS simulation based on classical electrodynamics does not reproduce this effect. Interestingly, the disappearance of the HO mode can be explained by taking into account nonlocal response in the silver nanoparticles,\cite{Mortensen:2014} which gives rise to size-dependent damping of SPs,\cite{Kreibig:1969} for one thing. Nonlocal response is a consequence of accounting for the wave-vector dependence of the dielectric function (i.e., spatial dispersion), which, in a hydrodynamic approach,\cite{Raza:2015b} originates from convective and diffusive currents of the free-electron gas in metals.\cite{Mortensen:2014} In Supplementary Note 1 we explain in detail the absence of the HO resonance in the EELS spectra of the smallest particles, based on two assumptions: First, the silver nanoparticles are fully embedded in a homogeneous background environment, a valid assumption for the smallest nanoparticles ($R<10$~nm). The second assumption is that the non-classical behavior of the nanoparticles is well described by the generalized nonlocal optical response (GNOR) model,\cite{Mortensen:2014} which shows that size-dependent damping decreases the EELS signal from the HO modes significantly for particle radii smaller than approximately 4~nm. The size-dependent damping is theoretically expected to increase with multipolar order $l$ and is therefore anticipated to influence the HO modes more than the dipole mode.\cite{Raza:2015b} The EELS simulations based on the GNOR model accurately explains our experimental observation of the disappearance of HO modes for $R<4$~nm, which provides strong support for the interpretation that this effect is due to nonlocal response, i.e., a fundamental physical property of silver nanoparticles, and not, e.g., due to the energy resolution of EELS or other instrument-related effects.

The second measured feature is a huge, nearly discontinuous, increase in dipole resonance energy of approximately 0.9~eV when the particle radius decreases from 4~nm to 1~nm (see Fig.~\ref{fig:fig5}), which is not reproduced in our classical simulations, see also Supplementary Note 1 for additional simulations with a different impact parameter. We are unable to explain the observed blueshift (shown in Figure 5) as a consequence of the position of the electron beam, since the measured resonance energies are still larger than the HO mode energy but lower than the bulk plasmon energy. Here, we note that the blueshift of the SP in similar-sized silver nanoparticles has been observed before.\cite{Kreibig:1985,Kreibig:1995,Ouyang:1992,Tiggesbaumker:1993,Charle:1998,Scholl:2012,Raza:2013} Previous EELS measurements of silver nanoparticles on thin ($<10$~nm) TEM substrates\cite{Scholl:2012,Raza:2013} have shown a (continuous) blueshift of the SP of approximately 0.5~eV in the same size range, which suggests that the non-classical blueshift depends on the effective background permittivity, with larger permittivity giving rise to larger blueshift. However, these previous EELS measurements were performed by acquiring spectra with impact parameters at the particle surface, suggesting, in the light of the results shown in this work, that HO modes may have come into play as well, thereby smoothening the blueshift of the SP to appear more continuous. Calculations based on the nonlocal GNOR model provide qualitative agreement with the EELS measurements shown in Fig.~\ref{fig:fig5}, however, the experimentally-measured blueshift is larger than predicted by this nonlocal theory (see Supplementary Note 1). The origin of the blueshift has been attributed to several other effects besides nonlocal response,\cite{Ruppin:1973,Raza:2013,Mortensen:2014} including quantum-size effects,\cite{Scholl:2012,Genzel:1975} screening from the \textit{d}-band electrons in silver,\cite{Liebsch:1993} and the combination of the latter two.\cite{Monreal:2013} In this work we have shown that nonlocal response indeed affects the SP modes of silver nanoparticles, but as the particle radii decreases, it becomes inherently more difficult to distinguish the relative importance of each of these effects as they all contribute non-negligibly on this size scale. This suggests that the origin of the observed blueshift likely derives from a concerto of these effects.

\section{Discussion and conclusion}
Using EELS, we have experimentally studied the plasmonic response of silver nanoparticles with radii $1-20$~nm encapsulated in a conformal silicon nitride coating. In our EELS spectra, we find, depending on the impact parameter, up to two SP-related resonance peaks due to the excitation of the dipole and higher-order (HO) surface plasmon modes, providing the first unambiguous observation of HO modes in nanoparticles with radius down to only $4$~nm. We find that impact parameters close to the surface of the nanoparticles exhibit the strongest EELS signal from the HO modes, while impact parameters several nanometers outside of the surface mainly probe the dipole mode. For impact parameters close to the inner surface of  nanoparticles with certain radii, the contribution of HO modes to the EELS signal even exceeds that of the dipole mode. Comparison with EELS simulations based on classical electrodynamic theory provides excellent agreement down to particle radii of only $4$~nm. The continued agreement even down to these size-scales is an impressive testament to the qualities of classical electrodynamic theory.

The strong presence of HO modes in small silver nanoparticles has profound influence on the interpretation of previous EELS measurements on nanoparticles on substrates.\cite{Ouyang:1992,Scholl:2012,Raza:2013} In particular, we have provided thorough experimental evidence that the EELS signal depends strongly on the impact parameter. Positioning the electron beam as close as possible to the nanoparticle surface will excite the HO modes as well as the dipole mode, so that the single or second observed surface-plasmon peaks in the EELS spectra actually are to be understood as the combined effect of many multipole surface plasmon resonances. To ensure that the EELS signal stems only from the dipole mode, we propose to position the electron beam several nanometers from the particle surface. Hereby, signal from HO modes and the bulk plasmon may be minimized.

For particle radii below 4~nm, we observe the disappearance of the HO modes and an increase in dipole resonance energy with decreasing particle size. The former is a consequence of size- and $l$-dependent surface plasmon damping, which are both effects due to nonlocal response and thereby a fundamental physical property of silver nanoparticles. Regardless of the impact parameter, we find only a single resonance for radii below 4~nm, which increases in energy with decreasing particle size. The lack of dependence on impact parameter suggests that these very small particles no longer support separate bulk and surface modes, but only a single plasmonic mode. We interpret the nature of this mode as a surface dipole mode, since the mode can be excited with impact parameters outside the particle, a feature not possible for bulk plasmons. The dipolar nature of the plasmon mode is further substantiated by earlier optical far-field measurements on similar-sized nanoparticles,\cite{Kreibig:1995,Tiggesbaumker:1993,Charle:1998} which only probe the dipole mode.

\appendix*
\section{Methods}
\subsection{Fabrication}
The silver nanoparticles are produced by a gas aggregation magnetron sputtering cluster source (NC200U-B, Oxford Applied Research Ltd.) in DC mode using different powers (20-30~W), argon flows (15-20 sccm) and aggregation distances (20-80~mm)\cite{Haberland:1994,deHeer:1993,Wegner:2006} to obtain a large size distribution on a single sample. For some depositions helium/hydrogen (10\%) is added (10 sccm) to stimulate the aggregation of smaller particles. The background and operation pressures are set to $2 \times 10^{-8}$~mbar and $1.7 \times 10^{-3}$~mbar, respectively, with a total deposition time of approximately 20~s. The silver sputter target has a purity of 99.99\%.

The silicon nitride thin films are obtained by magnetron sputtering (AJA International, Inc.) with 120~W and 40~sccm Ar flow of a Si$_3$N$_4$ target (99.9\% purity). The background and operation pressures are set to $2 \times 10^{-7}$~mbar and $1 \times 10^{-2}$~mbar, respectively. Two nominally 15~nm silicon nitride thin films are used to sandwich the silver particles. The sample is kept under vacuum during the entire fabrication process. The SiN$_x$-silver particle composite layers are deposited on silicon nitride TEM grids (5 nm thickness) with 9 windows (SiMPore, SN100-A05Q33A).

\subsection{EELS measurements}
The EELS measurements are performed with a FEI Titan transmission electron microscope (TEM) equipped with a monochromator and a probe aberration corrector. The microscope is operated in scanning TEM (STEM) mode at an acceleration voltage of 120~kV, providing a probe diameter of 0.5~nm and a zero-loss peak width of $0.15 \pm 0.05$~eV. The probe diameter is used as an estimate for the error in the impact parameters $b$. The EELS spectra are recorded with acquisition times ranging from $80$~ms to $150$~ms. To further improve the signal-to-noise ratio we accumulated up to 10 spectra for each measurement point.

The EELS spectra are analyzed by first removing the positive tail of the zero-loss peak using a power-law fit in the energy interval $1 - 2$~eV. To remove the background contribution of the EELS signal, a linear fit over a narrow energy range below the plasmon peaks and a broader range above the plasmon peaks is performed and subtracted from the spectrum. For large impact parameters ($b > R$), the EELS data contains primarily signal from the dipole resonances, which is fitted to a single Gaussian function using a nonlinear least-squares fit. For impact parameters close to the particle surface ($b \approx R$), the EELS data show a HO peak along with the dipole peak, to which we fit the sum of two Gaussian functions. The resonance energies are extracted from the Gaussian fit and the error in the resonance energies is given by the 95\% confidence interval for the estimate of the position of the center of the Gaussian function.

\subsection{Image analysis}
The analysis of the STEM images of the embedded nanoparticles is performed using the Image Processing Toolbox in MATLAB. We model the spatial extent of the STEM probe as a Gaussian function with a full-width at half-maximum of 0.5~nm, which we subsequently deconvolve from the image using the Lucy--Richardson deconvolution algorithm to sharpen the image. The sharpened grayscale image is converted to a black-white (binary) image using the threshold determined from Otsu's method. Subsequently, we accurately determine the boundary of the particle by a series of morphological operations. In particular, we clean the image by removing isolated pixels, fill out isolated interior pixels, and then erode and dilate the image using a disk-shaped structuring element (see Supplementary Note~4 for further details). Finally, the boundary of the particle is extracted and fitted to a circle and ellipse. The radius $R$ of the fitted circle is taken as the particle's radius, while the difference between the major $a$ and minor $b$ axes of the ellipse is used as an error bar for the radius (i.e., $\Delta R = a - b$).

\subsection{Simulations}
The EELS simulations are performed with the MNPBEM toolbox for MATLAB,\cite{Hohenester:2012} which solves the fully-retarded Maxwell's equations in the presence of a swiftly-moving electron using the boundary element method (BEM).\cite{Abajo:2002} We model the icosahedrally-shaped silver particles as spheres.\cite{Noguez:2007,Yang:2011} The permittivity for silver is taken from Ref.~\onlinecite{Johnson:1972}. We model the precise geometry of silver nanoparticles encapsulated in finite-thickness silicon nitride layers (see Supplementary Note 1 for a sketch of the theoretical model and further details). The value for the permittivity of the silicon nitride layer, $\varepsilon_{\text{SiN}_x}=3.2$, is obtained by fitting the simulated dipole resonance energies to the experimentally-measured dipole resonance energies. In Figures~\ref{fig:fig2}, \ref{fig:fig3}, and \ref{fig:fig5} the simulated EELS spectra have been convoluted with a Lorentzian function with a FWHM of 0.15~eV.

\begin{acknowledgments}
We would like to thank Ulrich Hohenester for his assistance with the MNPBEM Toolbox. The Center for Nanostructured Graphene is sponsored by the Danish National Research Foundation, Project DNRF58. The A. P. M{\o}ller and Chastine Mc-Kinney M{\o}ller Foundation is gratefully acknowledged for the contribution toward the establishment of the Center for Electron Nanoscopy. N.~S. acknowledges financial support from the Lundbeck Foundation (Grant No. R95-A10663). N.~A.~M. and M.~W. acknowledge funding from the Danish Council for Independent Research, (Grant No. 1323-00087). This work has been supported by a Marie Curie Career Integration Grant (M.~D.~V.), Project No: 293687.
\end{acknowledgments}

\bibliography{Raza_embedded_np_bib}

\end{document}